\documentclass[preprint]{aastex}

\shorttitle{Stellar Populations in the Antennae}
\shortauthors{Kassin, Frogel, Sellgren, Tiede, \& Pogge}

\slugcomment{\hbox{AJ Accepted [13 June 2003]}}

\begin{document}

\title{Stellar Populations in NGC 4038/39 (The Antennae): Exploring
A Galaxy Merger Pixel-by-Pixel\altaffilmark{1}}

\altaffiltext{1}{Based in part on observations obtained at the
Cerro Tololo Interamerican Observatory, operated by the Association of
Universities for Research in Astronomy, Incorporated, under a
cooperative agreement with the National Science Foundation.}

\author{Susan A. Kassin\altaffilmark{2},
Jay A. Frogel\altaffilmark{2,3,4},
Richard W. Pogge\altaffilmark{2,4},
Glenn P. Tiede\altaffilmark{4,5},
K. Sellgren\altaffilmark{2,4}
}

\altaffiltext{2}{Department of Astronomy, Ohio State University,
140 W. 18th Ave., Columbus, OH 43210-1173, kassin@astronomy.ohio-state.edu,
frogel@astronomy.ohio-state.edu, pogge@astronomy.ohio-state.edu,
sellgren@astronomy.ohio-state.edu}

\altaffiltext{3}{Presently at NASA Headquarters, 300 E. Street SW,
Washington, DC}

\altaffiltext{4}{Visiting Astronomer, Cerro Tololo Interamerican Observatory}

\altaffiltext{5}{Department of Astronomy, University of Florida,
Gainesville, FL, 32611, tiede@astro.ufl.edu}

\begin{abstract}

We present deep, photometrically calibrated BVRJHK images of the nearby
interacting galaxy pair NGC 4038/39 (``The Antennae'').  
Color maps of the images are derived, and those using the B, V, and K-bands
are analyzed with techniques developed for
examining the colors of stars.  From these data we derive
pixel-by-pixel maps of the
distributions of stellar populations and dust extinction for the
galaxies.  Analysis of the stellar population
map reveals two distinct episodes of recent star formation: one currently in
progress and a second that occurred $\sim$600 Myr ago.  
A roughly 15 Gyr-old population is found which traces the old disks of the galaxies
and the bulge of NGC 4038. The models used successfully reproduce the locations of clusters, and
the ages we
derive are consistent with those found from previous {\it Hubble Space
Telescope} observations of individual star clusters.  We also
find 5 luminous ``super star clusters'' in our K-band images that
do not appear in the B or V-band images.  These clusters are located in the overlap
region between the two galaxies, and are hidden by dust with visual
extinctions of $A_V\ga3$\,mag.  The techniques we describe in this paper should
be generally applicable to the study of stellar populations in galaxies
for which detailed spatial resolution with {\it Hubble} is
not possible.

\end{abstract}


\section{Introduction}
\label{sec:level1}
The Antennae is a well-known example of a strongly interacting pair of 
nearby spiral galaxies (at 19.2 
Mpc assuming $H_o=75$ $\rm km^{-1} s^{-1} Mpc^{-1}$; \citealt{W99}) viewed nearly face-on. With this favorable
geometry, we can examine the stellar content in the disks of these 
galaxies without complications arising from inclination or heavy dust obscuration, as is often 
the case for more advanced mergers (\citealt{sand}, and references therein). The 
galaxies in the Antennae are merging slowly on wide orbits (e.g.,
\citealt{miho}). Thus, we would expect multiple episodes of enhanced star
formation,  
triggered during times of closest approach of the two systems, to be separated far enough in 
time from each other to be observed as separate entities in the
merging galaxies.
 
The fact that enhanced episodes of star-formation accompany mergers of galaxies has 
been hinted at for almost 50 years, at least since \citet{zwic} noted
that the filaments connecting 
multiple galaxies are mostly blue.  However, even though mergers may play a dominant role in 
determining the morphology of galaxies, recent evidence has shown that the total amount of 
star-formation over the course of a merger is relatively modest compared to the amount of star-formation 
over a galaxy's lifetime \citep{kenn}. For example, the strength of the burst star-formation rate 
in NGC 4038/9 resulting from their recent merger is at most a few times that of a steadily star-
forming disk \citep{kenn}. By studying the Antennae in detail, we
will not only gain insight into the 
nature of the newly formed populations, but we will also learn about the roles of old, 
intermediate-age, and young populations in NGC 4038/9, and the distribution of dust in the merger. 

Optical and near-infrared colors provide a natural way to sort out stellar populations, as 
shown by \citet{frog}. Near-infrared colors are less sensitive to recent star-formation than 
optical colors and are relatively insensitive to the severe extinction present at optical 
wavelengths; the extinction optical depth at K is only 10 percent of
that in the V-band \citep{mart}. Since mergers are often dusty systems, the contribution of the K-band data to our 
analysis plays a crucial role in sorting out the amount of dust present, and thus helps us 
determine the stellar populations present. Near-infrared observations will be most 
sensitive to radiation from the old stellar population of the Antennae, i.e. those stars present 
before the most recent bursts of star-formation. For nearly all galaxies the old population 
constitutes most of a galaxy's stellar mass. Optical colors, on the other hand, are most sensitive 
to young Population I blue stars and will be most affected by dust. Even though the young stars 
dominate the optical light from a galaxy, they account for only a small fraction of the total mass 
of a galaxy. 

The Antennae have been studied in great detail with the Hubble Space Telescope (HST); 
these studies have concerned themselves primarily with discrete sources in the two galaxies 
(i.e., star clusters). Many ``super star clusters'' have been identified in the Antennae.  These 
stellar systems, often found in merging galaxies, present an opportunity to study the formation 
processes of globular clusters (e.g., \citealt{holt}). Extensive
work on the Antennae by \citet{W95}, 
\citet{W99} (hereafter, W99), and \citet{W01} (hereafter, ZFW) has concentrated 
on these super star clusters, in addition to less extreme clusters.
These authors find that the clusters range in age from very 
young $<5$ Myr-old clusters, which must have formed in the most recent merger event, to 
very old $\sim12-15$ Gyr-old clusters, probably true globular clusters which must have been present 
in the progenitor galaxies before the interaction. All told, they identified 14,000 point-like 
objects, with approximately 8000 of them being young clusters (using an
estimate based on subtracting all definite stars). 
       
The super star clusters, however, tell only part of the story of
star-formation in the Antennae.
Our purpose in this paper is primarily to analyze the more extended
stellar populations throughout the galaxies, to which 
clusters may or may not be contributors, such as the disks and 
bulges of the galaxies.
This gives a more complete picture of the star-formation in the
merger.  Although our data consist of
BVRJHK ground based images of the Antennae, we make use of 
only the B, V, and K band data; the near-infrared K band being
crucial in our population analysis.  An especially effective way to study the diffuse 
light of galaxies (e.g., \citealt{abra}, \citealt{eskr}, \citealt{kong}), or
a merger \citep{degr}, is to examine images pixel by pixel, thus 
exploiting the full spatial resolution of the images.    Using spectral 
synthesis of the observed colors in each pixel, we find the luminosity-weighted
average population(s) present, and assign a 
luminosity-weighted average age to each pixel.  From this, we derive effective
population age and reddening maps of the Antennae that we use to
study the locations of young, intermediate-age, and old populations, along 
with the dust distribution.  In $\S 2$ we 
describe the observations and data; in $\S 3$ we explain the
techniques used to obtain ages
and reddening.  The derived age and extinction maps, in addition to the
clusters detected only at K, are examined in $\S 4$; our
conclusions are given in $\S 5$.

\label{sec:level2}
\section{Observations}

B, V, and R images of the Antennae were obtained under photometric conditions on UTC 
1995 March 8 on the 1.5 m telescope of the Cerro Tololo Interamerican Observatory (CTIO) with 
the direct imaging filter wheel and the $1024\times1024$ Tek No.2
CCD. This CCD gives a field of $7\arcmin 5 \times 7\arcmin 5$
with a spatial scale of 0\farcs44\,pixel$^{-1}$. Cumulative on-source 
integration times were 20 min, 15 min, and 10 min at B, V, and R, respectively. Twilight sky 
exposures were used to provide flat-field calibration. The images in each filter were added 
together and cosmic rays were cleaned using interactive median filtering. Equatorial standard 
star fields from \citet{land} were observed at a range of airmasses to derive photometric 
transformations onto the Kron-Cousins BVR system. The final images
have limiting surface 
brightnesses ($3 \sigma$ rms) of 24.6, 24.4, and 23.8 mag arcsec$^{-2}$ for B, V, and
R, respectively. 

J, H, and K images of the central regions of the Antennae were obtained through thin 
cirrus on UTC 1995 March 13 with the CTIO 1.5 m telescope using the CTIO Infrared Imager 
(CIRIM). Each of the images covered a field of view of $\sim5\arcmin \times5 
\arcmin$  with a spatial scale of 1\farcs16\,pixel$^{-1}$.  
The J and H images were taken in groups of 5 equal length integrations
with comparable length 
off-target sky images acquired between integrations. Among the 5 images, the target 
position was dithered on the detector by $30\arcsec$ in a cross-shaped pattern to allow for 
median combination and to avoid bad pixels and detector blemishes. The K images were acquired as 
10 separate images taken in two sequences of 5 images each using the same sky chopping and 
target dithering pattern as the J and H images. The integration time
of each image that went into each of the 10 K-band images is only 5
seconds.  We would chop to sky each time we took a 5 second image.  Each sequence of images in a given filter began 
and ended with off-target sky images of comparable exposure to the on-target images, 
providing a complete set of sky frames for subtraction. Cumulative on source exposure times 
were 30, 23, and 43 minutes at J, H, and K, respectively. 

Because the long exposure images were acquired under non-photometric conditions, we 
also took short J, H, and K images of the Antennae under photometric conditions two nights 
later. Standard stars from the list of \citet{cart} were observed at a range of 
airmasses on the night the short images were taken. The long exposure images were calibrated 
using stars common to both long- and short-exposure images. Photometry was reduced to the 
CTIO/CIT system \citep{elia}. The final images have limiting
surface brightnesses ($3 \sigma$ rms) of 21.4, 20.8, and 19.7 mag
arcsec$^{-2}$ for J, H, and K, respectively.

To facilitate comparison between images in different photometric bands, we merged the 
IR and optical data sets into 6 photometrically calibrated images on a
common pixel scale of 1\farcs16\,pixel$^{-1}$, corresponding to a
linear scale of 108 parsecs per pixel.  Our final images have a 
modest seeing of $1\farcs5$ FWHM (140 parsecs effective linear
resolution).  We 
used the near-IR images as the fiducial ones since they have the largest pixels. The optical 
images were placed on the scale of the near-IR ones with linear transformations that included 
translation, scaling, and rotation terms.  These transformations were derived by fitting to the 
positions of stars common to the images. Inter-image registration is accurate to $\pm0\farcs2$, as 
estimated from the variances in the centroids of field stars in the final registered images. These 
images, in which each pixel in each image samples the same location in the galaxy, form the 
basic data of our subsequent analysis. As examples, Figures 1a, 1b, and 1c display the B, V, and K 
images, respectively.  Absolute uncertainties in the photometry are
estimated to be $\pm 0.02$ mag at B, 0.03 at V, and 0.06 at K. The
larger uncertainties in the near-infrared reflect our resort to a
``bootstrap'' calibration between photometric and non-photometric images.

\section{Analysis}

The main objective of this paper is to determine the stellar content of the Antennae as a 
function of position.  We are especially interested in deriving
luminosity-weighted ages for the different populations.  
As mentioned in the Introduction, our approach to this objective will
be based on a pixel-by-pixel 
analysis of the images.  Pixels are excluded from the analysis if they are below a threshold 
value of 5$\sigma$ above the median sky value at K (18.8 mag/ pixel;
19.2 mag/ arcsec$^2$).  This criterion is adopted since contribution from the 
sky is greatest in the K-band and because the K image is the noisiest of the three.   Since all 
of our images are aligned and transformed onto the same pixel scale,
pixels rejected from the K-band image are also rejected from the other bands. 

We model pixels from the Antennae with two single stellar population models: 
one which models exponentially decaying star-formation on a timescale
of $\tau$= 100 Myr, the typical dynamical time for a galactic disk, 
and another with a decay timescale of $\tau$= 15 
Gyr, which describes an older but still star-forming disk, from the 1996 version of the Bruzual $\&$ 
Charlot GISSEL models (\citealt{bruz}; hereafter, BC). The models have a Salpeter 
initial mass function truncated at 0.1 and 100 M$_{\odot}$ and solar
metallicity.  This is in agreement with 
\citet{meng} who found that the spectra of a number of clusters
in the Antennae was consistent with solar metallicity. 
We cannot rule out the possibility, however, that there was steady
enrichment of the stars over the last 15 Gyr, which would result in
older stars that are more metal-poor.  Nonetheless, the qualitative results of this
paper do not differ if a $\tau$= 15 Gyr model with sub-solar metallicity
is used.

The colors $V-K$ and $B-V$ were chosen because 
$V-K$ has a long wavelength baseline that gives good leverage for
measuring dust extinction, and $B-V$ is a strong indicator of the 
presence of a young population (young clusters form an age sequence in
increasing $B-V$).  Color maps for $B-V$ and $V-K$ are 
in Figures 2a and 2b.  Young clusters (white) are the most apparent feature 
on the $B-V$ image, while $V-K$ reveals a large amount of dust 
(black), in addition to some unobscured clusters (white).

In Figures 3a and 3b we plot contours of pixel densities 
for $V-K$ versus $B-V$ from the
5$\sigma$ cuts of NGC 4038 and 4039, respectively.  These pixels have
not been dereddened.
BC models are plotted in Figures 3a and 3b as thin and thick lines for the 
decay times of 100 Myr and 15 Gyr, respectively.  
Contours are for pixel densities of 10-70 pixels in steps of 20, with the 
dotted contours tracing the outer envelopes of the distributions of pixels.
The contours for pixels from NGC 4038
are also plotted in Figure 3c, and ages are marked on
the models in this figure in units of Gyr.  Reddening vectors are
plotted for $\rm A_V$ of 1.0 in all panels of Figure 3.

We estimate the age of each pixel from the 5$\sigma$ cuts in Figures
3a and 3b by dereddening the pixels to have the 
same colors as the stellar population models, using the reddening law
given by \citet{card} with an $\rm R_V$ of 3.1. 
We assume that the light in each pixel is dominated by 
a single stellar population and that the offset of a pixel from the model locus is due entirely to 
reddening. The age of the location on the stellar population model that the pixel dereddens to is 
assigned to the pixel, and the dereddening necessary to move the pixel to the model line is 
noted. The goal is to produce both age and reddening maps of the Antennae.
By doing this analysis pixel-by-pixel, we not only analyze the populations found in star clusters 
(as in previous analyses), but we also find ages of diffuse light
present in the images, where the contribution from clusters is variable.
 
Figures 4a and 4b show histograms of the number of dereddened pixels with different ages for NGC 
4038 and 4039, respectively, modeled with the $\tau$= 15 Gyr model. Both histograms show a 
tall narrow peak at $\sim15$ Gyr along with a smaller peak at a much younger age of less 
than one or two Gyr.  We associate the pixels in the 15 Gyr peak with the old underlying disk 
and bulge populations of the galaxies and we take the younger peak to be evidence for a recent 
starburst population.  
Figure 5 is an image of the stellar ages in the Antennae, where pixels from
the 15 Gyr peaks of Figures 4a and 4b are shaded black. We 
interpret these pixels as composing the older star-forming disks and bulges of the Antennae, 
and assign ages to them from the model with a decay time of 15 Gyr. These pixels follow 
the distribution of W99's old globular clusters.

We model the remaining pixels (excluding the 15 Gyr pixels) with
the $\tau$=100 Myr model.  
Age histograms for these pixels are shown in Figures 4c and 
4d for NGC 4038 and 4039, respectively. 
We are insensitive to ages $\ga 1$ Gyr, where reddening vectors are
almost exactly parallel to the $\tau$=100 Myr model.  Reddening
vectors are not parallel to the $\tau$=100 Myr model for
ages $\la 1$ Gyr.  Figure 4c shows 3 age peaks
in NGC 4038: around 90, 300, and 600 Myr. Poissonian error bars are
plotted for each bin in the histogram.  These 3 peaks appear to
be significant.  When we examine where in
the images the pixels in each peak lie, we find the following.
Pixels from the first peak outline the young star clusters.  
Those from the second peak 
tend to be found near pixels from the first peak, and thus also outline the
young star clusters.  However, pixels from the third peak tend to be found
toward the outer regions of NGC 4038's spiral arm and 
some in the region of overlap between the two
galaxies.  Pixels from the youngest peak in NGC 4039 (Figure 4d) have
a median age of $\sim$ 400 Myr.  When we look for their location
in the images, we find that they are primarily situated in 
the bulge and spiral arm, with a few pixels located along the outskirts of the
disk.  Those in the small peak at $\sim$ 800 Myr in NGC 4039 are 
most likely due to noise; these pixels are sparsely 
scattered along the faint outskirts of the disk of NGC 4039 and are 
most often isolated pixels.
In Figure 5, pixels are shaded according to age, with the oldest ones 
shaded the darkest.
Note that errors in photon counting statistics are about $1\%$, and 
calibration errors are $2\%$ for B, $3\%$ for V, and $6\%$ for K. The
typical color error bar plotted in 
Figure 3a implies an error in age that is approximately our bin size
in the logarithmic age histograms in Figures 4c and 4d (0.1 dex).

Are the separate 90, 300, and 600 Myr peaks of Figure 4c and the
400 Myr peak of Figure 4d artifacts?  If they are,
then we would expect to detect random errors in the 
ages of these pixels, for example, finding
young pixels in regions where we would expect to find intermediate-aged or old pixels.
Since we observe age gradients in the 90 and 300 Myr pixels, 
which is a systematic trend, the pixels in these peaks
cannot be due to random noise.
A similar argument applies to pixels from the 400 Myr peak in Figure
4d and the 600 Myr peak in Figure 4c.  The 400 and 600 Myr pixels are
spatially separated from the other peaks and trace structure 
in NGC 4039 (spiral arm and bulge) and in NGC
4038 (outer part of the spiral arm), respectively.
Finally, as we shall detail in $\S 4.2$, the ages of the peaks
are consistent with previous observations.

We have thus far assumed that the dereddened colors in each pixel map
onto a unique age in the single stellar population models.  For those
pixels that map onto an intermediate-age population (90, 300, and 600
Myr for NGC 4038, and 400 Myr for NGC 4039; Figures 4c and
4d), their colors might plausibly be due to a mix
of young ($\lesssim40$ Myr) and old ($\sim 15 $ Gyr) stars.  One way
to address this issue is to use the properties of a linear color, or
``flux-ratio,'' diagram. Linear color diagrams were used to identify the components of composite 
stellar populations by \citet{rabi} and \citet{frog}. With the aid of
a linear color diagram, we find that the colors of pixels in the
$\sim 600$ Myr peak can only be due to an intermediate-age single stellar population; no reasonable
combination of young and old stars with the models we have used here 
can produce such colors.
In Figure 5, pixels from the $\sim 600$ Myr peak in NGC 4038's
histogram (Figure 4c) are 
shaded darker than those of the younger peaks.  They are  
spatially distinct from the young ($\lesssim40$ Myr) population
and are located in the outer parts of NGC 4038's disk and
the overlap region between the two galaxies. The
$\sim 600$ Myr pixels 
trace the intermediate-aged ($\sim500$ Myr) clusters reported by W99.
On the other hand, we find that pixels in the 90 and 300 Myr peaks of NGC
4038 and the 400 Myr peak of NGC 4039 can
be due to a mixture of old and young starlight.  These pixels tend to lie in
regions surrounding the young clusters in Figure 5.

Unfortunately, with our data, stellar population models
have little diagnostic power for ages younger than 
40 Myr.  We therefore take the pixels that  
do not deredden to the model line (i.e., those that deredden below the
model line in Figures 3a and 3b) and assign ages to them of 
$\lesssim40$ Myr.  We call these ``zero-age'' pixels since the models have no
diagnostic power below 40 Myr.  On the age map in Figure 5 they compose the 
white regions and clearly trace the 
sites of the youngest clusters. These pixels are coincident with most of the
$\lesssim 30$ Myr clusters in W99 and are found in regions with H$\alpha$ emission.
Another reason why these pixels deredden below the model line may be 
due to added contribution to the broadband flux by a gaseous
emission component (see \citealt{char} and references therein,
\citealt{zack}, \citet{ande}, and \citealt{krue}).

\section{Morphology}
\subsection{Age and Extinction Maps}

The luminosity-weighted age map of the Antennae in Figure 5 shows that large-scale morphological features 
correlate with age. 
The youngest pixels (those which do not deredden to the model line and 
which are shaded white) are found in the centers of clusters in the very dusty ``Northeastern 
Star Formation Region'' (Region I; $A_V$ $\sim2-3$), the ``Western Loop''
(Region II; $A_V$ $\sim1.5-2.5$), and 
the ``Northwest Extension'' (Region III; $A_V$ $\sim$1.5-2.5), using
nomenclature from Figure 5a of W99. Our K-
band image (Figure 1c) aids in the detection of this young population, especially in Region I, 
since its light can penetrate much of the dust present. Notice the many high surface brightness 
clusters in Region I, along with patches of dust in the $V-K$ image in Figure 2b. An age gradient 
is detected surrounding the star-forming clusters in the regions
mentioned above, which may also be due to the mixing of light from the old disk population
with a contribution from the zero-age clusters that decreases as
pixels are located with increasing distance from 
the young clusters. This gradient is apparent in Figure 5 as the shading darkens with 
increasing distance from the centers of clusters.
Current star-formation is also readily apparent 
in 4 small knots surrounding the nucleus of NGC 4038.  Young
pixels of ages $\sim60$ Myr are found near their centers, and
pixels with ages of up to $\sim350$ Myr are found
on the outskirts of the knots.  \citet{meng1} find knots near the 4038 nucleus to
be $65\pm15$ Myr old from K-band spectra.  The nucleus of NGC 4039 has an age of 
$\sim55$ Myr at its center (\citealt{meng1} measure $65\pm15$ Myr for
it from K-band spectra), knots in the arm emerging from the nucleus are about 80-90 
Myr, and ages of about 100-400 Myr are found in the diffuse light surrounding the knots in 
the arm. These populations can also be explained as a combination of old and young
populations. All these star-forming regions are apparent in both the B and
the V-band images in Figures 1a and 1b and the $B-V$ color map in Figure
2a. W99 calculate ages of $\sim5-10$ Myr for 
clusters in Region II, one third of clusters with ages $\sim100$ Myr
and two thirds with ages $\lesssim$ 30 
Myr with ``streaks'' of recent star-formation in Region I, less than $\lesssim5$ Myr for the young clusters 
in the ``Overlap Region,'' and ``somewhat older'' in the arm of NGC 4039 from a mix of GHRS 
spectroscopy, H$\alpha$, and UBVI colors from HST.  
The youngest pixels follow the H$\alpha$ map in ZFW.
ZFW find that the H$\alpha$ flux is not followed by highly reddened young clusters,
as expected, but is instead closely associated with $\la10$ Myr young bright clusters.

We detect a $\sim600$ Myr intermediate-aged population in the diffuse light in an arc to the west 
of Region II, to the east of Region I, to the north of the ``Northwest Extension,'' to the south 
of the NGC 4038 nucleus, and on the east side of the ``Overlap
Region.'' The $\sim600$ Myr 
population appears to be a remnant of NGC 4038's 
pre-encounter star-forming disk. W99 identify intermediate-age clusters ($\sim500$ Myr old), ``with 
the most obvious members located in the `Western Loop'" (Region III). Our $\sim600$ Myr old 
population follows these clusters. W99 comment that these clusters appear to have formed in a 
separate burst during a previous encounter of the two galaxies. 

In Figure 5, the old $\sim 15$ Gyr underlying disk and bulge
population of NGC 4038
and 4039 is shaded black.  This population permeates the disk of NGC
4039 and is prominent only in the inner parts and outskirts of NGC
4038's disk.  

In Figure 6, we present an extinction map of the Antennae where areas shaded black 
represent the dustiest regions and have $A_V$ of $\sim2-3$. White areas demarcate the zero-age 
clusters that do not deredden to a model locus. There is a huge mass of dust apparent in the 
``Overlap Region'' ($A_V \sim2-3$). Dust clouds of varying thickness
($A_V$ $\simeq1.5-2.5$) are detected 
throughout the disk of NGC 4038. The dust lane passing through the
nucleus of NGC 4038 has $A_V$ $\sim2.0-2.8$.
Most of the diffuse light in this disk has an $A_V$ of $\sim1$. The 
spread in $A_V$ for pixels from NGC 4038 is evidenced by the spread of
pixels in $V-K$ in Figure 2b. NGC 4039's nucleus and small spiral arm, each with an extinction of $\sim2.4$ and 
$\sim1.2-1.6$, respectively, are quite dusty. The inner disk of NGC 4039 does not contain many dust 
clouds and has an $A_V$ of $\sim1$, similar to NGC 4038's inner
disk. This is well illustrated in Figure 2a 
where the pixels from NGC 4039 do not have as great of a spread in $V-K$ as those from NGC 
4038.  

A low-resolution CO map shows 3 major concentrations: the NGC 4038
nucleus, the ``Overlap Region,'' and the NGC 4039 nucleus (ZFW).  A
high-resolution CO map shows a more extended distribution in the ``Overlap
Region'' and Region II (ZFW).  ZFW also 
find that highly reddened young star clusters
correlate well with molecular clouds.  Most of the
gas in the merger is found in the ``Overlap Region,'' along with more than half of
all highly reddened clusters (ZFW).  They also find bright young
clusters to generally be near peaks of the CO emission.
As seen in the age map in Figure 5, our youngest clusters do indeed 
come from these highly reddened regions.  ZFW find that young
bright clusters ($\la 10$ Myr) have a mean $A_V$ of $\sim1.5$, and that
clusters with ages $\ga 100$ Myr (but still younger than the
intermediate-age population) have a mean $A_V$ of $\sim0.3$.  These
$A_V$ values are consistent with our extinction map.  
We also find reddened clusters in the ``Overlap Region'' in our
K-image which are not detected at B or V.
In short, the far- and mid-infrared maps presented in
ZFW resemble the CO maps, which in turn resemble the extinction
map in Figure 6.

\subsection{``Hidden'' Clusters Present in the K-band}

We have detected 5 objects on our K-band image for which we obtain
only upper limits in the B and V-bands, and identify them as heavily 
reddened star clusters.  Some have been previously studied with 
spectroscopy, as discussed below. Their locations are indicated with open 
circles on the B and K-band images in Figure 7, with
clusters numbered from 1 to 5 on the K-band image in Figure 7a.  
All are located in the very dusty ($A_V$ $\sim2-3.2$) 
``Overlap Region'' between the two galaxies.  
K-band magnitudes of these clusters were measured 
in a square aperture, 3 pixels (3\farcs48) on a side, 
corresponding to $\sim 320$ parsecs.  
Given our seeing of $\sim$1\farcs5 FWHM at K, these could 
in fact be multiple (2 or 3) clusters unresolved by our seeing,
as it is well known that massive clusters in the Antennae
are often clustered (ZWF). Upper limits ($3\sigma$)
for B, V, $B-K$, and $V-K$ are given in Table 1.  The uncertainty in 
B, V, $B-K$, and $V-K$ is 0.1, while the uncertainty in K is 0.06.

Clusters 2 and 5 have near-infrared magnitudes reported in the
 literature that are also listed in Table 1.
\citet{meng1} observed cluster 2, also known as cluster 80 from \citet{W95}, 
with integral field K-band spectroscopy.  They derived a K-band 
magnitude from the
spectroscopy in a 2\farcs2 box (205 parsecs), and measured a V-band magnitude
from W99's V-band image in the same aperture.
\citet{gilb} also imaged cluster 2 in the K-band, and quoted a V-band
 measurement from the HST image.  They determined a half-light radius of
$\sim32$ parsecs and found that the cluster releases strong
 far-ultraviolet flux that excites the surrounding medium on scales
 of up to 200 parsecs. While the K-band measurements for cluster 2 
from our measurements, \citet{meng1}, and \citet{gilb} are consistent 
considering the apertures used, V-band measurements differ
 between authors.  Because of our modest seeing ($1\farcs5$), 
we do not directly detect cluster 2.  Instead, we
 measure an upper limit of $21.2\pm 0.1$.  This is consistent with
 \citet{gilb} who obtain V=23.5 (via a private communication with B.C.
 Whitmore $\&$ Q. Zhang 1999 who measure it from the HST image).
 \citet{meng1} measure V=$18.8\pm0.15$ for cluster 2, which is not
 consistent with either our measurement or that of \citet{gilb}.
Cluster 5 is resolved into at least 2 separate clusters on
the W99 image, cluster 355 from \citet{W95} and cluster 15 from W99.
These 2 parts were observed by \citet{meng}, and they give magnitudes 
for V and K$_{\rm short}$ in a 1\farcs74 radius.  In Table 1 we sum
 \citet{meng}'s measurements for these 2 clusters.
Our measurements of this cluster most likely differ from \citet{meng} 
because of a combination of factors.  Namely, the seeing on our image is
 large, we use a different aperture than \citet{meng}, and we calculate
 sky counts locally, in an area which could be contaminated by
 light from neighboring clusters.  In addition, \citet{meng} noted (see
 the ``Comments'' column in Table 1 of their paper), that there were
 difficulties keeping both targets in the slit during integrations.

 \citet{meng1} derive an $A_V$ for cluster 2 of 4.3 from the
flux ratio of Br$\gamma$/H$\alpha$.
\citet{gilb} estimate the screen extinction
to the cluster to be $A_V=9-10$.   The extinction values found from
these higher resolution observations are larger than those determined from
our data since, with a higher resolution, one can see in between dust features 
which are smoothed over in our images.
\section{Conclusions}

With simple modeling, we have mapped the distributions of stellar populations and dust in the
Antennae, and derived the star-formation history 
of the encounter between NGC 4038 and 
4039. This was done with a pixel-by-pixel analysis of modest-seeing (1\farcs5) 
ground-based images.

We detected two epochs of star-formation that presumably resulted from the two most 
recent close encounters of the galaxies. In brighter regions, where we do not expect a significant 
contribution from an underlying older ($\sim15$ Gyr) population, we discerned a very young, 
essentially zero-age burst.  We called it ``zero-age'' because we have no diagnostic power 
with single stellar population models for very young populations where, below 40 Myr, colors
are degenerate with reddening for our data.  
Pixels of ``zero-age'' correspond to the brightest blue star-forming regions 
associated with the HII regions seen by W99. An intermediate-aged ($\sim600$ Myr) population was 
found to lie in the outer regions of NGC 4038's spiral arm; 
it is the same population as the one detected in 
W99's HST study.  Populations with ages between about 40 and 500 Myr can be
explained as either single or composite populations, depending on
assumptions about reddening.  These populations outline young
star-forming regions in the Antennae.  The intermediate-age ($\sim600$
Myr) population
most likely formed in a burst during the first encounter 
between the two galaxies which resulted in the tidal tails. Current star-formation in the 
Antennae appears to be confined to locations where there was previous star-formation activity, 
as evidenced by the location of the intermediate-aged population,
along the NGC 4038 spiral arm. In addition to the young and intermediate-aged 
populations, we detected an old star-forming disk population of $\sim15$
Gyr which traces the disk of NGC 4039 and the inner disk and
bulge of NGC 4038. 

Based on \citet{deva}'s classification of NGC 4038 as an SB(s)mpec and 
NGC 4039 as an SA(s)mpec, we suggest that NGC 4038 was probably a gas-rich Sb/Sc galaxy 
before the encounter, while NGC 4039 was probably a relatively gas-poor Sa. The encounter 
between the two that occurred about 600 Myr ago could have consumed most of the gas in 
NGC 4039.  Thus, within this galaxy we found very few ``zero-age'' clusters. NGC 4038, being of 
later type, still retained a significant amount of its copious gas content after the first 
encounter. It is this gas that is responsible for the current ``zero-age'' burst of star-formation and 
that gives NGC 4038 its dramatic appearance.  We infer that the product of a merger between 
an Sa and an Sc galaxy, especially if observed at a higher redshift, would probably resemble a 
disturbed Sc galaxy if viewed in a rest-frame optical band at a similar stage in a merger 
sequence as the Antennae. This is due to the prominence of star-formation in the disk of NGC 
4038, compared to activity elsewhere in the merger, and to the low surface brightness of other 
features such as tidal tails and the meager amount of star-formation in NGC 4039.  

Our extinction map delineates where dust accumulates in the merger. While there is a 
large dust cloud where the two galaxies overlap, much less extinction is detected over most of 
the galaxies' disks.  This is expected, since in a merger interstellar gas and dust are destabilized 
and fall toward the central potential well.  In addition to reddening, our K-band 
image allows us to find buried star-formation regions that are not
detected in our B or V-band images, clearly showing the value of
near-infrared imaging of dusty mergers.

The success of the pixel by pixel analysis when applied to a complex system such as the 
Antennae suggests that this simple method could be used to study 
the stellar content and star-formation histories of galaxies for which comparably detailed HST 
imagery is not available. The key 
is to have the rest-frame H or K-band in addition to rest-frame visible colors to create a long 
baseline in wavelength that reaches to the near-infrared.  This
baseline is needed to gain leverage with the dust 
content and thus be able to do the population analysis, and, at least in this case, to reveal a few 
surprises (i.e., the buried clusters).  

\acknowledgements
We thank the CTIO TAC for generous allocation of time for the OSU
Galaxy Survey, and the anonymous referee for their thorough and
thoughtful comments. 
Funding for the OSU Bright Spiral Galaxy Survey was provided by grants from The National 
Science Foundation (grants AST-9217716 and AST-9617006), with additional funding by the 
Ohio State University. This research has made use of the NASA/IPAC Extragalactic Database 
(NED) that is operated by the Jet Propulsion Laboratory, California Institute of Technology, 
under contract with the National Aeronautics and Space Administration.

\clearpage
\begin{figure}
\caption{B, V, and K-band images of the Antennae are in
  (a), (b), and (c), respectively.  NGC 4038 is the north-westernmost galaxy (top), and
 NGC 4039 is to the southeast (bottom).} 
\end{figure}

\clearpage
\begin{figure}
\caption{$B-V$ and $V-K$ color maps of the
  Antennae. $V-K$ reveals a large amount of dust (black), in addition to
  unobscured clusters (white). The most apparent feature on the $B-V$ image
  is the young clusters (white).}
\end{figure}

\clearpage
\begin{figure}
\epsscale{1}
\plotone{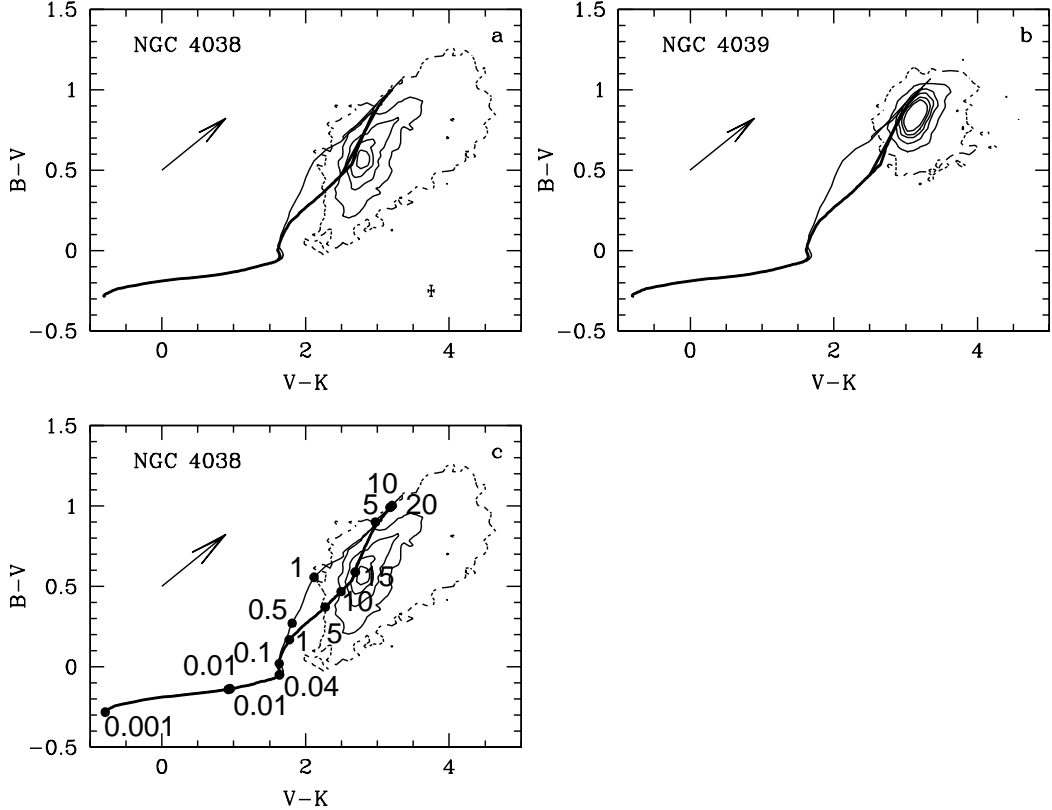}
\caption{Color-color diagrams for pixels from images of
  NGC 4038 and 4039 which have not yet been dereddened are plotted in
  (a) and (b), respectively. 
   Contours are drawn for pixel densities of 10-70 pixels in 
   steps of 20; the dotted contours trace the outer envelopes of the
   distributions of pixels. Stellar population models for decay times of 
  $\tau=100$ Myr and $\tau=15$ Gyr are plotted as thin and thick
  lines, respectively, in (a) and (b). In (c) these models are labeled with ages
  in Gyr, and pixel density contours from NGC 4038 are plotted.
Reddening vectors are drawn for $A_V$ of 1.0, and
   an error bar for the calculation of each pixel's dereddened color is
   plotted in (a). We assume that the light in each pixel is
  dominated by a single stellar population and that the offset
  from the model locus is due entirely to reddening. 
  We estimate the age of the pixels
  by dereddening each to have the same colors as the stellar
  population models. 
The age of the
  location on the stellar population model that the pixel dereddens
  to and the dereddening necessary to move
  the pixel to the model line is assigned to each pixel to create age
  and reddening maps of the Antennae.}
\end{figure}

\clearpage
\begin{figure}
\epsscale{1}
\plotone{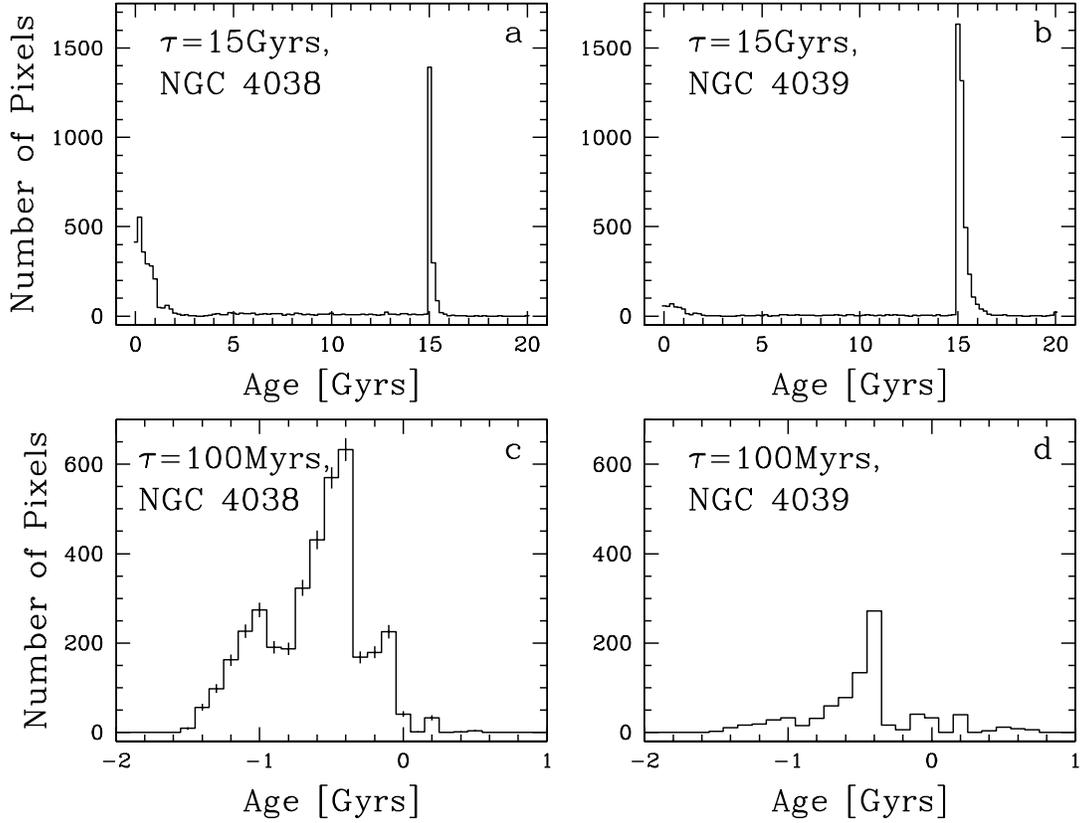}
\caption{Age histograms  
  created by dereddening pixels from NGC 4038 and 4039 to a BC
  model with a decay time of $\tau$= 15 Gyr 
  are in (a) and (b), respectively. 
  We associate the pixels in the 15 Gyr peaks of (a) and (b) with the old underlying
  disk and bulge populations of the galaxies and take the younger
  peaks to be evidence for a recent starburst population.
  Age histograms created by dereddening pixels that are not in the 15
  Gyr peaks of (a) and (b)  
  to a BC model with a decay time of $\tau$= 100 Myr are in (c) and (d), respectively.
  Populations in (c) and (d) with ages between about 40 and 500 Myr can be explained
  as either single or composite populations.  These populations
  outline the young star-forming regions in the Antennae. Pixels in
  the $\sim$ 600 Myr peak of (c) can only be due to an intermediate-age
  single stellar population; no reasonable combination of young and
  old stars can produce such colors. Poissonian error bars are plotted
  for each bin in (c).}
\end{figure}

\clearpage
\begin{figure}
\epsscale{0.75}
\plotone{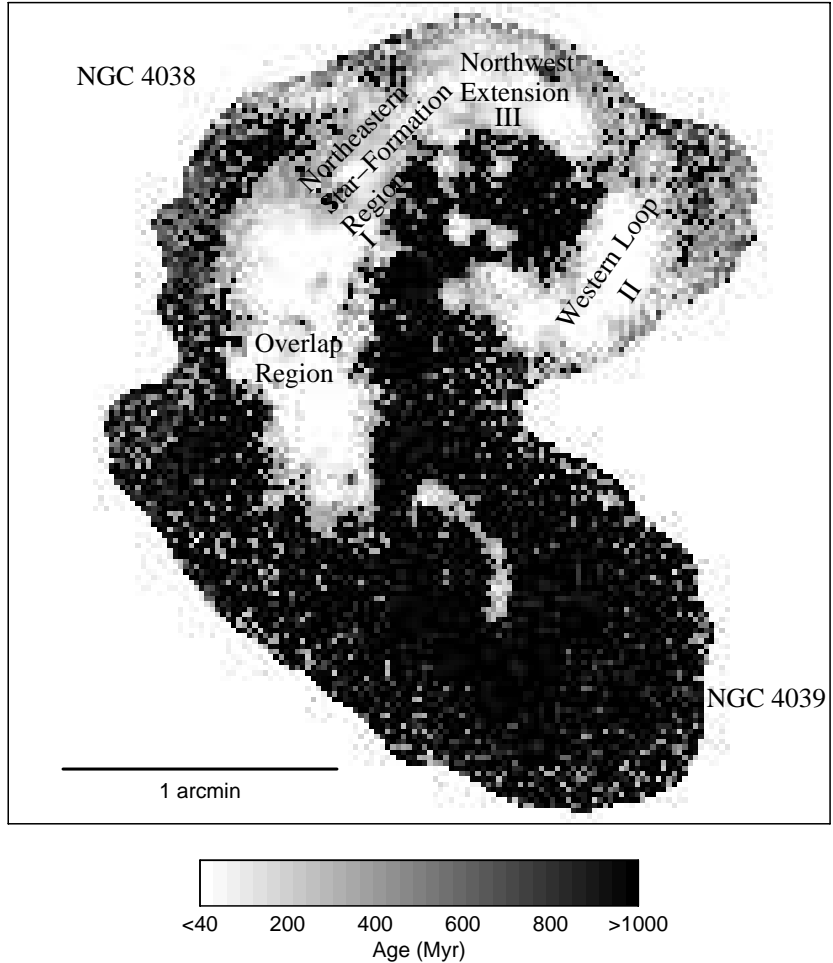}
\caption{Age image of the Antennae created by
  dereddening pixels in Figures 3a and 3b to the $\tau = 15$ Gyr and
  $\tau = 100$ Myr models.  This image shows that
  large-scale morphological features correlate with age. The youngest
  pixels are shaded white and are found in the centers of the
  clusters.  
  An age gradient is detected surrounding these young clusters, 
  which can also be due to mixing of
  light from the old disk population with a contribution from the
  young clusters that decreases as pixels are located with increasing distance
  from the young clusters. A $\sim 600$ Myr intermediate-aged
  population, found mostly in the outer parts of NGC 4038's spiral arm
  with some in the ``Overlap region,'' is spatially 
  distinct  from the young clusters.  An old
  underlying disk population permeates NGC 4039 and
  is found only in the very inner and outer parts of NGC 4038's disk.}
\end{figure}

\clearpage
\begin{figure}
\epsscale{0.75}
\plotone{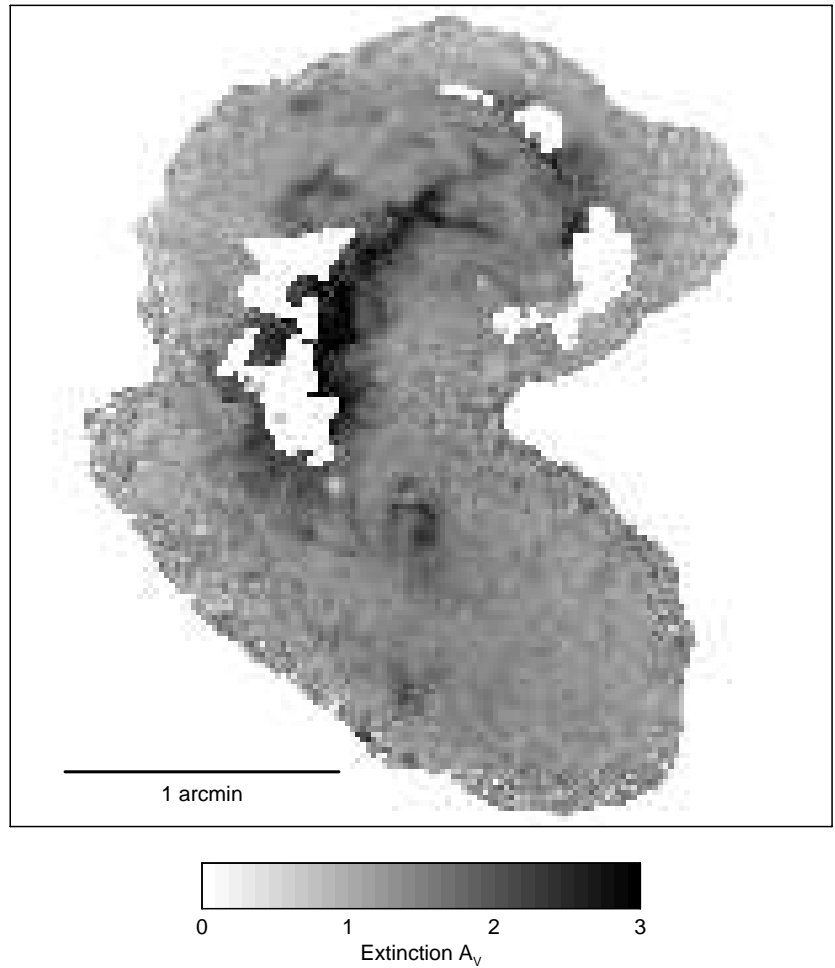}
\caption{Dust map of the Antennae where areas shaded
  black represent the dustiest regions and have $A_V \sim 2-3$.
  White areas demarcate the youngest clusters that do not deredden to
  a model locus.  There is a huge mass of dust apparent in the overlap
  region between the two galaxies, and dust clouds of varying thickness
  are seen through the disk of NGC 4038.  Most of the diffuse light in
  NGC 4038's disk has an $A_V \sim 1$.  NGC 4039's nucleus and small spiral arm, each with an
  extinction of $\sim 2.4$ and $\sim 1.2-1.6$, respectively, are quite
  dusty.  The inner disk of NGC 4039 does not contain many dust clouds
  and has an $A_V$ of $\sim 1$, similar to NGC 4038's inner disk.}
\end{figure}

\clearpage
\begin{figure}
\caption{Heavily reddened star clusters detected at K for which we obtain
  only upper limits at B and V. The left panel indicates
  their locations with open circles on the K-band image; the right
  panel shows the locations of these clusters on the 
B-band image. All are located in the very dusty $A_V \sim 2-3.2$
  ``Overlap Region'' between the two galaxies.  K-band magnitudes and
  upper limits on the $B-K$ and $V-K$ colors for a $3 \sigma$
  detection are given in Table 1.}
\end{figure}

\clearpage
\begin{deluxetable}{cccccccccccc} 
\tabletypesize{\small}
\tablecolumns{12} 
\tablewidth{0pc} 
\tablecaption{``Hidden'' Clusters Present in the K-band} 
\tablehead{ 

\colhead{} &\multicolumn{5}{c}{This paper\tablenotemark{a}}
&\multicolumn{3}{c}{Gilbert et al.} 
&\multicolumn{3}{c}{Mengel et al.}\\

\cline{2-12}\\ 

\colhead{cluster}
&\colhead{B} &\colhead{V} &\colhead{K}
&\colhead{$B-K$} &\colhead{$V-K$}
&\colhead{V} &\colhead{K}
&\colhead{$V-K$} 
&\colhead{V} &\colhead{K}
&\colhead{$V-K$}
}

\startdata 
1 &$>21.2$ &$>20.6$ &16.33 &$>4.9$ &$>4.3$ 
&\nodata &\nodata &\nodata 
&\nodata &\nodata &\nodata\\

2 &$>21.7$ &$>21.2$ &14.85 &$>6.9$ &$>6.4$ 
&23.5\tablenotemark{b} &14.6\tablenotemark{b} &8.9\tablenotemark{b} 
&18.8\tablenotemark{c} &14.8\tablenotemark{c} &4.0\tablenotemark{c}\\

3 &$>22.1$ &$>21.2$ &14.94 &$>7.2$ &$>6.3$ 
&\nodata &\nodata &\nodata 
&\nodata &\nodata &\nodata\\

4 &$>21.5$ &$>20.4$ &16.04 &$>5.5$ &$>4.4$ 
&\nodata &\nodata &\nodata 
&\nodata &\nodata &\nodata\\

5 &$>19.3$ &$>21.6$ &15.60 &$>3.7$ &$>6.0$ 
&\nodata &\nodata &\nodata 
&19.31\tablenotemark{d} &15.04\tablenotemark{d} &4.27\tablenotemark{d}\\

\enddata 
\tablenotetext{a}{Measured in a 3\farcs48 box.
 Uncertainties are 0.1, for B, V, $B-K$, and $V-K$. 
 K has an uncertainty of 0.06.}
\tablenotetext{b}{The half-light radius is measured to be
 $\sim$ 32 parsecs. The $V-K$ color is calculated from the 
 HST V-band image.}
\tablenotetext{c}{From \citet{meng1}.  Measured in a 2\farcs2 box. K-band is measured from spectroscopy; V-band is from W99 image.}
\tablenotetext{d}{From \citet{meng}.  The sum of measurements of cluster 15 from 
W99 and 355 from \citet{W95}, each measured in a 1\farcs74 radius. K
is K$_{\rm short}$, and the V-band is measured from the W99 image.}
\end{deluxetable} 

\end{document}